# Dynamics of He++ ions at interplanetary and Earth's bow shocks

Olga V. Sapunova[1,*], Natalia L. Borodkova[1], Georgii N. Zastenker[1], Yuri I. Yermolaev[1]

[1] Space Research Institute of the Russian Academy of Sciences (IKI RAS),117997, Moscow, Russia
  nlbor@mail.ru (N.L.B.); gzastenk@iki.rssi.ru (G.N.Z.); yermol@cosmos.ru (Y.I.Y.)
* Correspondence: sapunova_olga@mail.ru;

**Abstract:** Experimental investigations of the fine plasma structure of interplanetary shocks are extremely difficult to conduct due to their small thickness and high speed relative to the spacecraft. We study the variations in the parameters of twice-ionized helium ions (4He++ ions or α-particles) in the solar wind plasma during the passage of interplanetary shocks and Earth's bow shock. We use data with high time resolution gathered by the BMSW (Bright Monitor of Solar Wind) instrument installed on the SPEKTR-R satellite, which operated between August 2011 and 2019. The MHD parameters of He++ ions (the bulk velocity $V\alpha$, temperature $T\alpha$, absolute density $N\alpha$, and helium abundance $N\alpha/Np$) are analyzed for 20 interplanetary shocks and compared with similar parameters for 25 Earth's bow shock crossings. Measurements from the WIND, Cluster and THEMIS satellites were also analyzed. The correlations in the changes in helium abundance $N\alpha/Np$ with the parameters $\beta i$, $\theta Bn$ and MMS were investigated. A correlation between $N\alpha/Np$ and the angle $\theta Bn$ was found: the lower the value of $\theta Bn$, the greater the drop in helium abundance $N\alpha/Np$ falls behind the IP shock front. For Earth's bow shock crossings, we found a significant increase in the helium abundance $N\alpha/Np$ in quasi-perpendicular events.

## 1. Introduction

Collisionless shocks remain one of the most mysterious processes of space plasma [1]. To form a shock, the speed of the incident plasma flow must exceed the speed of the obstacle by the value of the magneto-sound speed. Two types of shock waves were considered in this experimental study: Earth's bow shock and interplanetary (IP) shocks. Compared to an Earth shock, the velocity of an IP shock relative to a satellite is several times higher. There are few qualitative measurements with sufficiently high temporal resolution for the plasma parameters at the IP shock front, and these are of great scientific interest. IP shocks (including those generated by so-called high-speed streams from coronal holes and fast coronal mass ejections) propagating in the solar wind are one of the main agents for the transfer of perturbations from the Sun to the Earth [e.g., 2-6]. All of the parameters of the solar wind plasma, such as the velocity, temperature, density, and magnetic field, change dramatically as the IP shock front passes. It is well-known that the two main ionic components of the solar wind are protons and twice-ionized helium ions (He++ or α-particles). The ratio of single-ionized helium ions in the solar wind does not exceed a few thousandths of the density of He++ ions [7], and is beyond the scope of this paper.

Variations in the proton and He++ ion parameters and the $N\alpha/Np$ at large distances of $> 10^6$ km are directly related to the properties of the upper corona of the Sun and the mechanisms of solar wind formation in this region. It is therefore important to determine the relative density of helium relative to the variations in the main (proton) component due to local physical processes at small distances of $\sim 10^3$ km [8-13 and references therein].

Studies of IP shocks and the time profiles of various components of the solar wind during propagation are therefore inter-related tasks associated with the same problem. The changes in He++ ions at the IP shock fronts and their interactions were first studied by Gosling et al. [14], after which this issue was investigated via both modeling [e.g. 15-17] and experimental means [e.g. 2-4]. Most of the experimental results in this area have been obtained from magnetic field and proton measurements. However, the majority of space plasma instruments do not have sufficiently high time resolution for the investigation of He++ ion parameters in the fine structure in the immediate vicinity of the ramp.

These data became available after the launch of the SPEKTR-R satellite (apogee - 333 570 km, and perigee - 576 km) in 2011 and the operation of the BMSW (Bright Monitor of Solar Wind) instrument as part of the PLASMA-F experiment onboard the satellite. This instrument allowed researchers to study the fine structure of the IP shock front with high time resolutions of 0.031 s for the magnitude and direction of the solar wind ion flux, and 1 s for the velocity, temperature and ion density (for protons and for He++ ions) [e.g. 18-22] scales of IP shock front are identical and about equal to that of the magnetic field; the ion scales are determined by the proton thermal gyroradius; the numerous fast oscillations of the ion flux occur near the front ramp and the majority of observed variations of helium abundance is generated by turbulence.

The main aim of this brief article is to examine the variations in the density of the He++ ions at the front of an IP shock and to detect changes in the parameters of these ions directly next to the ramp with a high time resolution. A similar analysis is carried out for Earth's bow shock, and the results for both types of shocks are compared.

In the following sections of the article, we explain our selection of data sources and compare measurements of the plasma and magnetic fields from different satellites. We give examples of IP shock front crossings with the relevant solar wind plasma parameters, including variations in the He++ ion parameters, analyze the changes in $N\alpha/Np$ during the passage of the IP shock front for various values of $\beta i$, $\theta Bn$ and MMS, and discuss our results.

## 2. Materials and Methods

### 2.1 BMSW data

In this study, we used data obtained from the BMSW plasma spectrometer, which operated between August 2011 and the beginning of 2019. BMSW was intended for measurements of the energy spectra of ions in the range 0.2–2.8 keV/Q, proton bulk velocities in the range 200–750 km/s, an isotropic ion temperature of 1–100 eV, and an ion density of 1–100 cm−3. The time resolution of the BMSW instrument was 0.031 s for the magnitude and direction of the ion flux, and 1 s for the velocity, temperature and density of the protons and He++ ions. In some cases, an instrument mode was also available that allowed for measurements of the velocity, temperature and density of protons with a resolution of 0.031 s. Detailed descriptions have been presented in several prior papers [e.g. 18-21]. The methodical limitations of this approach are discussed in Section 3.3.

### 2.2 Additional data

Although a magnetometer was installed onboard the SPEKTR-R satellite, it was not operational due to technical problems from the very beginning. To investigate the time profiles of the main solar wind plasma parameters, we supplemented the BMSW data with measurements of the magnetic field (with a time resolution of 0.092 s) from the MFI magnetometer on the WIND spacecraft [23], which was located in the solar wind near the first libration point. In some cases, magnetic field data from the THEMIS-B/THEMIS-C [24] or Cluster 1-4 [25] satellites were analyzed. We also used data obtained with the 3DP instrument [26] on the WIND spacecraft with a time resolution of 3 s (excluding February - May 2012 when 3DP was being re-calibrated) to compare the time profiles for the protons and He++ ions of the solar wind at different points in space.

### 2.3 Methods of data analysis

#### 2.3.1 Comparison of values obtained from different satellites

As described above, data from several satellites were used in this study, and we therefore needed to take into account the separation distance between the different spacecraft. For this reason, the shock fronts and the associated structures observed by WIND and SPEKTR-R may not be identical. Magnetic field and plasma fluctuations can occur in the solar wind. However, Weygand et al. [27] showed that it is valid to assume that the fluctuations in the IP magnetic field are 'frozen in' at distances equal to the L1 Lagrange point to the Earth, while over larger distances, this assumption first breaks down for the fast solar wind. The presence of magnetic field fluctuations in the slow solar wind therefore has little effect on its quasi-stationary structure. At the same time, it is known that IP shocks usually propagate in the slow solar wind, which has fluctuations in its density. Matthaeus et al. [28] investigated the spatiotemporal correlations of plasma turbulence in the solar wind and found that for a slow solar wind, the correlation persisted over a larger time separation period. There are also inhomogeneities along the front itself, meaning that the collisionless shock front structures measured by SPEKTR-R and WIND might differ noticeably from each other. According to Eselevich and Eselevich [29], the spatial scale at which the solar wind can be considered uniformly dense along the shock front in the ecliptic plane is about (4–8) • $10^6$ km, a fairly large distance. Since the properties of He++ ions are investigated at MHD scales, the magnetic field measurements made onboard WIND can be used for shock analysis.

To illustrate the above, Fig. 1 gives an example of the time profiles of the main solar wind plasma parameters and the magnetic field during the IP shock passage on 9 July 2017. It shows the solar wind density, temperature, and proton velocity registered by the SPEKTR-R spacecraft (panels a–c) and the magnitude and components of the magnetic field registered by the WIND spacecraft (panels d–e) that were time-shifted to the location of SPEKTR-R.

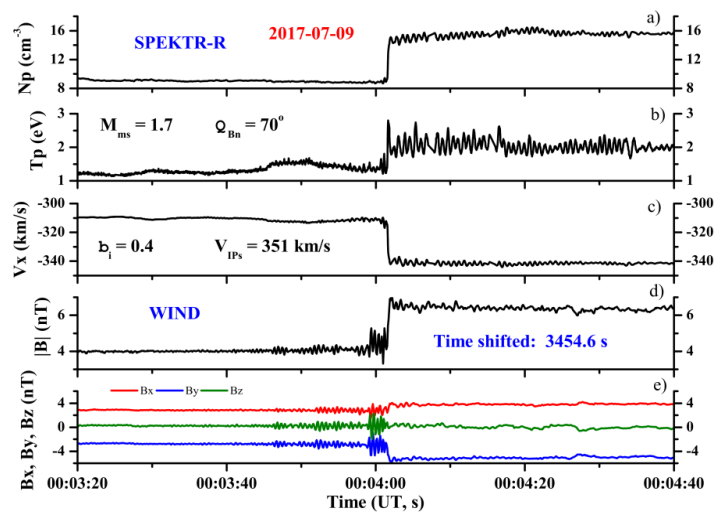

**Figure 1**. Time profiles of plasma parameters and magnetic field during the IP shock passage on 9 July 2017. From top to bottom: (**a**) Solar wind density; (**b**) Solar wind temperature; (**c**) Solar wind velocity (along the Earth–Sun direction) of the ion flow; (**d**) Magnitude of the magnetic field and (**e**) The components of the magnetic field.

From these profiles, we can observe the following structural features of the front: a ramp with duration of ~ 0.34–0.35 s, an upstream oscillation wave train with a wavelength of about ~ 0.37–0.47 s, and a downstream wave train with a duration of about ~ 0.64–0.84 s. This example shows good agreement between the values obtained from the two satellites for the duration of the IP shock ramp and the upstream and downstream wave train durations for the same IP shock, determined based on the parameters of the solar wind plasma and the magnetic field.

We also compare the parameters of the He++ ions for these two spacecraft. Figure 2 shows an comparison of the densities (absolute densities for the protons and He++ ions, and relative densities for the He++ ions) obtained by the SPEKTR-R and WIND satellites for the very first IP shock in our database, dated 9 September 2011.

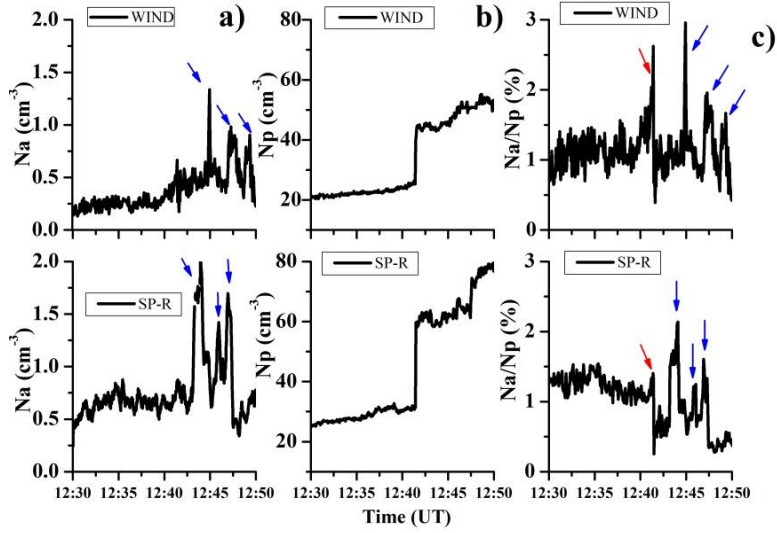

**Figure 2.** Changes in the time profiles of the proton and He++ ion densities according to data from the SPEKTR-R and WIND satellites for the event of 9 September 2011: (**column a**) Absolute He++ ion density; (**column b**) Absolute proton density; (**column c**) Relative He++ ion density.

The data from both instruments show structures in the density of the He++ ions. In column (a), the blue arrows indicate three structures with an increased absolute density of He++ ions. These are located downstream, and although their positions relative to the IP shock ramp are slightly different, the coincidence in the shapes and the numbers of peaks indicates good stability for the structures, given that we consider this a perturbed region. A certain difference in the absolute values of the proton density, which is noticeable in column (b), can be explained by the different sensitivity of the sensors of the instruments. However, it should be noted that the relative change in the proton density in the data from both instruments coincides, and has a value of Np2/Np1 = 1.85. The helium abundance Nα/Np is shown in column (c), and three structures can be observed which correspond to those in column (a). The red arrows indicate an increase in the abundance of helium Nα/Np immediately before the IP shock ramp, which is clearly visible from the data from both instruments, as well as a sharp decline after the ramp. Based on these matches, we can confirm that the use of data from different satellites at the scales considered here is acceptable.

*2.3.2 Calculation of characteristic shock parameters*

The bulk ion parameters (bulk velocities Vα and Vp, temperatures Tα and Tp, densities Nα and Np) were calculated using a standard procedure, assuming an isotropic Maxwellian distribution of ions over velocities. The basic shock parameters such as the magnetosonic Mach number $M_{ms}=V_{IP}/V_{ms}$ (where

$$V_{ms} = \left[0.5 * (V_A^2 + V_s^2) + (0.5 * (V_A^2 + V_s^2)^2 - V_A^2 * V_s^2 * cos\theta_{Bn})^{1/2}\right]^{1/2} \qquad (1)$$

and $C_A$ and $C_S$ are the Alfvénic and sound speeds), the angle $\theta_{Bn}$ between the magnetic field vector and the direction of the front normal, and the plasma beta

$$\beta_p = k * T_p * N_p/(B^2/8\pi) \qquad (2)$$

(the upstream kinetic-to-magnetic pressure ratio) were calculated for all IP and bow shock crossings. The normal to the front of IP shock $\bar{n}$ was determined using a geometric method based on data from four or more satellites. The estimation of $\bar{n}$ for the Earth's bow shock was made using the model developed by Verigin et al. [30] for the shape of the bow shock, based on data from nearby satellites, including an estimate of the direction of the magnetic field.

### 2.3.3 BMSW measurement of He++ ions

An example of the IP shock with a clearly visible flux of He++ ions is given in Fig. 3, which shows the energy time spectrogram of the solar wind ion flux registered by the BMSW instrument. Two populations can be clearly distinguished in the spectrogram, and these are indicated by arrows. Populations of both protons and He++ ions exist before, during and after the IP shock front crossing at 11:31 on 30 September 2012. Data from the BMSW instrument allow us to determine the distribution of the ions if the temperature and velocity of the protons are small [12].

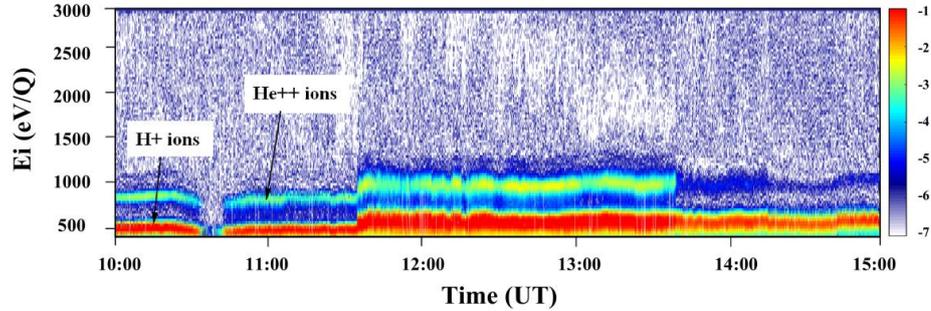

**Figure 3.** Energy–time spectrogram of the ion flux measured on 30 September 2012 by the BMSW instrument.

The dependence of the sensor current on the voltage shows two separate peaks at low ion temperatures: the proton peak (at ~500 V) and the He++ ion peak (at ~1000 V), as shown by the solid blue line in Fig. 4. The solar wind plasma accelerates and heats up behind the IP shock front, as a result of which the proton peak and the He++ ion peak expand and shift to the right of the sensor current–voltage curve. Since the range of energy measurements was limited to 3 keV, the second peak may be shifted beyond this range at high velocity.

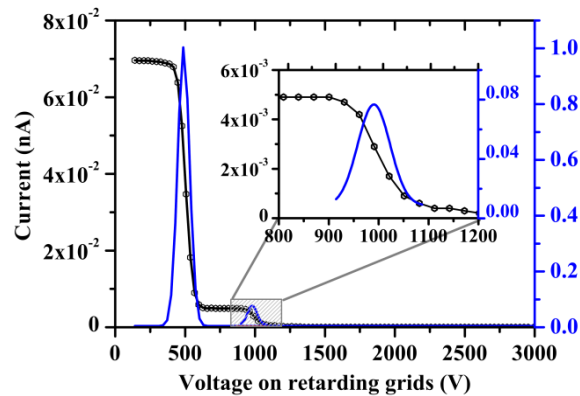

**Figure 4.** Example of a three-second spectrum from the BMSW instrument. The black line shows the characteristic energy cutoff, and the blue line shows the derivative of the FC currents for protons and He++ ions.

High temperatures ($T_p > 50$ eV) and velocities ($V_p > 600$ km/s) were the two main reasons why only a proportion of the 57 IP shocks were suitable for definition of the He++ parameters. We selected 20 IP shocks, and later 25 Earth bow shock crossings, for which it was possible to determine the flow of He++ ions during the passage of the front (Table 1). For each IP shock, the parameters of the protons and He++ ions were calculated: the bulk velocity ($V_p$, $V\alpha$), temperature ($T_p$, $T\alpha$), and density ($N_p$, and for He++ ions both $N\alpha$ and the abundance $N\alpha/N_p$). For the Earth's bow shock crossings, we calculated the densities $N_p$, $N\alpha$ and helium abundance $N\alpha/N_p$.

## 3. Results
### 3.1. Characteristics of shocks

A list of the events that were analyzed and the values of the characteristic parameters ($V_{IP}$, $\beta_p$, $\theta_{Bn}$, $M_{MS}$) for these events just before the shock ramp are presented in Table 1. The propagation velocity of the IP shock was quite low, and exceeded 500 km/s only in three events; the $M_{ms}$ mean ratio was 2.0, and hence was also not high. The parameter $\beta_p$ varied widely due to the different combinations of characteristics of the solar wind plasma and IP magnetic field. In the case of a low magnitude of the magnetic field (2–3 nT) and a relatively high proton density (10–15 cm$^{-3}$) and/or proton temperature (15–20 eV), $\beta_p$ could be larger than 2. In the first set of events (IP shock crossings), there was a lack of quasi-parallel cases ($\theta_{Bn} < 45°$), so this dataset was complemented by a second dataset (Earth's bow shock crossings). Na2(%)/Na1(%) is the change in the helium abundance $N\alpha/N_p$, i.e. the downstream-to-upstream ratio.

**Table 1.** Parameters for IP shock and Earth's bow shock crossings

| IP shock crossings | | | | | | Earth's bow shock crossings | | |
|---|---|---|---|---|---|---|---|---|
| Date | $V_{IP}$ | $\beta_p$ | $\theta_{Bn}$ | $M_{MS}$ | Na2(%)/Na1(%) | Date | $\theta_{Bn}$ | Na2(%)/Na1(%) |
| 09.09.2011 | 412 | 3.0 | 26 | 3.0 | 0.60 | 23.03.2012 | 77±4 | 7.3 |
| 01.11.2011 | 403 | 0.4 | 74 | 1.5 | 0.95 | 28.03.2012 | 39±4 | 1.8 |
| 15.05.2012 | 428 | 1.5 | 86 | 1.0 | 0.94 | 05.04.2012 | 71±5 | 3.5 |
| 21.05.2012 | 406 | 1.4 | 80 | 2.6 | 0.71 | 23.04.2012 | 52±2 | 4.5 |
| 03.09.2012 | 457 | 0.5 | 35 | 2.7 | 0.82 | 28.05.2012 | 21±2 | 3.1 |
| 30.09.2012 | 302 | 1.8 | 65 | 1.8 | 0.83 | 07.08.2012 | 81±5 | 5.1 |
| 08.10.2012 | 409 | 0.3 | 84 | 1.7 | 0.91 | 08.08.2012 | 65±5 | 9.5 |
| 13.04.2013 | 472 | 0.5 | 47 | 3.0 | 0.87 | 24.08.2012 | 85±5 | 11.0 |
| 23.04.2013 | 312 | 1.5 | 63 | 1.6 | 0.84 | 16.09.2012 | 45±4 | 4.1 |
| 18.05.2013 | 502 | 0.2 | 75 | 1.3 | 0.99 | 12.10.2012 | 88±4 | 7.2 |
| 19.04.2014 | 520 | 0.2 | 62 | 1.0 | 0.87 | 30.10.2012 | 76±5 | 4.4 |
| 03.05.2014 | 225 | 4.0 | 89 | 1.2 | 0.97 | 02.11.2012 | 29±5 | 2.0 |
| 07.06.2014 | 438 | 0.3 | 89 | 2.5 | 0.92 | 14.11.2012 | 84±4 | 5.6 |
| 03.07.2014 | 309 | 2.2 | 55 | 1.0 | 0.91 | 17.11.2012 | 64±3 | 7.3 |
| 17.03.2015 | 562 | 0.3 | 65 | 2.3 | 0.87 | 24.11.2012 | 83±2 | 5.0 |
| 21.06.2015 | 327 | 2.2 | 83 | 5.7 | 1.12 | 09.03.2013 | 84±5 | 6.9 |
| 12.10.2016 | 431 | 0.5 | 21 | 2.3 | 0.48 | 11.03.2013 | 80±4 | 4.2 |
| 09.11.2016 | 354 | 0.6 | 87 | 1.6 | 1.05 | 11.03.2013 | 60±4 | 2.6 |
| 31.08.2017 | 398 | 0.9 | 53 | 1.4 | 0.92 | 14.03.2013 | 76±4 | 7.9 |
| 21.10.2017 | 395 | 0.6 | 76 | 1.4 | 0.97 | 16.05.2013 | 22±5 | 5.3 |
| -- | -- | -- | -- | -- | -- | 09.06.2013 | 48±4 | 3.8 |
| -- | -- | -- | -- | -- | -- | 10.06.2013 | 27±4 | 3.5 |
| -- | -- | -- | -- | -- | -- | 06.07.2013 | 32±6 | 2.5 |
| -- | -- | -- | -- | -- | -- | 13.09.2013 | 50±4 | 7.7 |
| -- | -- | -- | -- | -- | -- | 01.12.2013 | 35±5 | 6.6 |

### 3.2. Variations in the He++ ion parameters during the passage of the IP shock

As the IP shock front passes, all of the parameters of the solar wind plasma change sharply and significantly. An example of this change is shown in Fig. 5 for the density, velocity, temperature, and magnetic field. Figures 5(a-c) show the plasma data from the SPEKTR-P and WIND satellite measurements. The proton parameters are shown in black for BMSW and in orange for the 3DP instrument, respectively. The blue line shows the parameters of the alpha particles based on data from the BMSW instrument. Figure 5(d) shows the magnitude and components of the magnetic field according to the MFI instrument on the WIND satellite. There are areas with a close match between the parameters, and also areas with some differences in values, but in general, the data from the two instruments give similar shapes. It should be noted that the proton bulk velocities match well, both upstream and downstream, and the slope of the IP ramp is the same for all of the proton parameters measured by different instruments. This confirms the stability of the IP shock ramp at MHD scales during propagation from WIND to SPEKTR-R.

There is a slight decrease in the helium abundance Nα/Np immediately after crossing the IP shock ramp. It then increases and levels out within approximately one minute. On average, the velocity of the He++ ions was slightly (about 7%) less than the velocity of the protons, both before and after the front, while the proton velocities recorded by the different instruments matched. The temperature of the He++ ions was twice as high as the temperature of the protons, due to the higher mass of α-particles, and rose significantly close to the IP shock ramp.

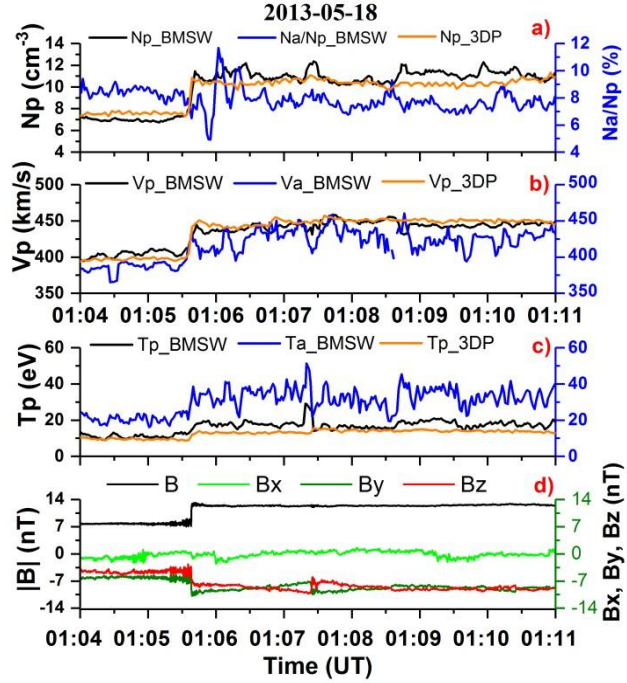

**Figure 5.** Time profiles for the solar wind plasma parameters and magnetic field during crossing of the IP shock on 18 May 2013: (**a**) densities; (**b**) velocities; (**c**) temperatures; (**d**) magnetic field magnitude and components.

To compare the helium abundance Nα/Np in the upstream and downstream regions, a statistical analysis was carried out. Figure 6 shows the distributions of Nα and Nα/Np for all events, before and after the passage of the front. It can be seen that after crossing the IP shock front, the absolute values change quite noticeably. In particular, the average value of the He++ ion density increases by a factor of about two. However, due to an equally strong increase in the density of protons behind the IP shock front, the Nα/Np changes only slightly.

For all these events, we investigated whether the changes in the helium abundance Nα/Np during the passage of the IP shock front depended on the values of $\beta_i$, $\theta_{Bn}$ and $M_{MS}$. The results are shown in Fig. 7.

It can be seen that there was no explicit dependence of the change in the relative density of He++ ions on the parameter $\beta_i$ (see Figure 7b). For both small and large values of $\beta_i$, a large or small change in the relative density of He++ ions was possible. There was also no clear correlation with the $M_{MS}$ parameter (see Fig. 7c). However, if we excluded one event with a large parameter value ($M_{MS}$ = 5.8), there was a tendency for the helium abundance Nα/Np to fall after the IP shock front as the $M_{MS}$ increased.

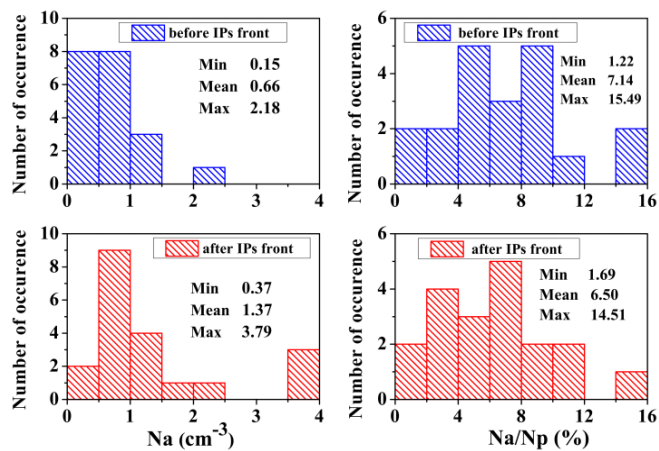

**Figure 6.** Statistical histograms of the absolute (**left**) and relative (**right**) He++ ion density in undisturbed and disturbed solar wind.

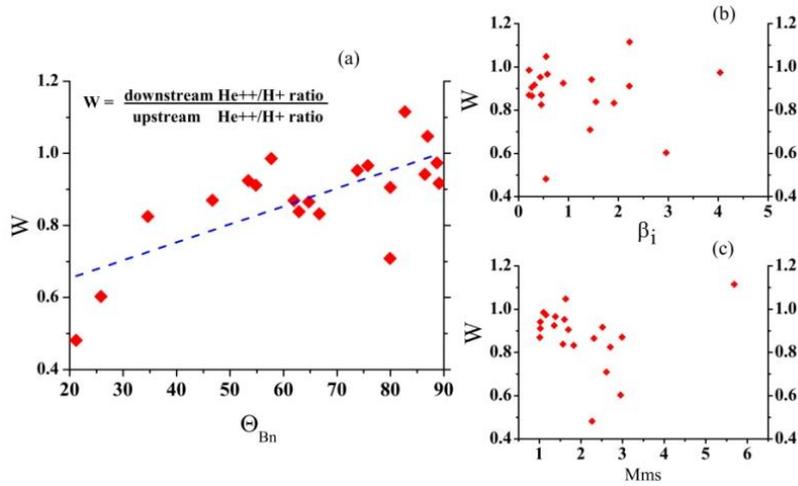

**Figure 7.** Dependence of the change in helium abundance Nα/Np during the passage of the IP shock front on the parameters (**a**) $\theta_{Bn}$, (**b**) $\beta_i$, and (**c**) $M_{MS}$.

However, we can see an obvious correlation between the change in the helium abundance Nα/Np and the angle $\theta_{Bn}$ (see Fig. 7a). The smaller the angle $\theta_{Bn}$, the greater the fall in the value. In other words, Nα/Np will fall much more (by a factor of 2–2.5) behind the front of a quasi-parallel IP shock than behind the front of a quasi-perpendicular IP shock.

Due to the small number of IP events, we increased the statistic, and processed data on the Earth's bow shock crossings gathered by the SPEKTR-R satellite in the period 2011–2013. Figure 8 shows the results, superimposed onto the previous dataset. The angle $\theta_{Bn}$ for these new bow shock crossing events was estimated using the model developed by Verigin et al. [30] for the shape of the bow shock and data from nearby satellites, including an estimate of the magnetic field direction. The new dataset allowed us to supplement the area of quasi-parallel events, with an angle of $\theta_{Bn} < 45°$. Despite the uncertainty in these estimates (shown in Fig. 8), there is a clear trend: the larger the angle $\theta_{Bn}$, the greater the increase in the helium abundance Nα/Np after the IP shock front. This trend coincides with that of the IP shock front crossings, as the helium abundance Nα/Np changes less in the quasi-parallel cases.

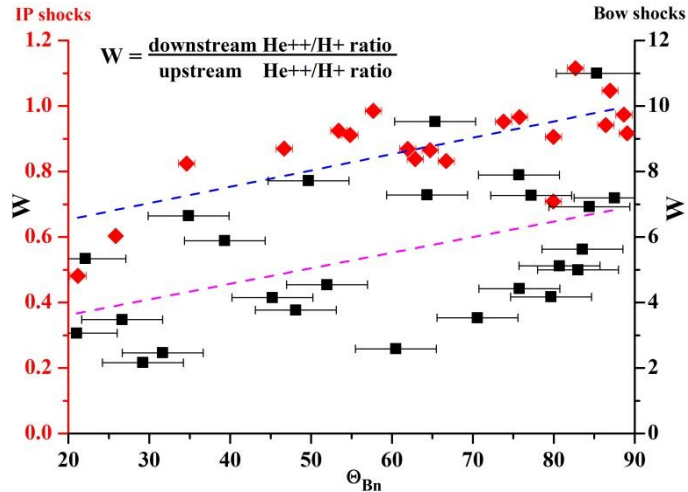

**Figure 8.** Dependence of the change in helium abundance Nα/Np on the angle $\theta_{Bn}$ at the intersection of the shock fronts. The red symbols show the events of the IP shock front crossings, and the black symbols show the Earth's bow shock front crossings. The blue dashed line shows the trend for the first set, and the purple dashed line shows the trend for the second set. The errors in the values of the angle $\theta_{Bn}$ are shown by bars.

However, a comparison of the two sets of events shows a significant difference in the values of the change in the helium abundance Nα/Np. For IP shock front crossings, the helium abundance Nα/Np usually becomes lower than in the unperturbed region, while for the Earth's bow shock crossing, this parameter always increases, in some cases by almost an order of magnitude.

## 4. Discussion

The drop in the helium abundance Nα/Np behind the IP shock and Earth's bow shock fronts during the transition from a quasi-perpendicular to a quasi-parallel case may be caused by the outflow of some ions from the disturbed region to the undisturbed one through the ramp due to a decrease in the $\theta_{Bn}$ angle, which makes this transition more efficient. In one study by Trattner and Scholer [17], the time profile of the reflected He++ ions was modeled for quasi-parallel shocks. The diffusion of a proportion of the ions (both protons and He++) into the undisturbed region was noted, and the relative density of He++ ions in the reflected stream can be compared to that of an undisturbed solar wind. Similar results have been reported for quasi-perpendicular cases [e.g. 31]. This result is consistent with the dependence that can be seen in Figs 7(a) and 8; in the case of a quasi-parallel shock, a significant proportion of the He++ ions can move to an undisturbed region, thus causing the relative density of the He++ ions behind the shock front to fall. Results published by Gosling et al. [14] suggest that the mechanism underlying the acceleration of low-energy ions by quasi-parallel shocks may be one-sided, which may also explain our results. For a more detailed study of this issue, it is necessary to increase the statistics of observation of quasi-parallel IP shocks.

In the case of a quasi-perpendicular shock, the value of Nα/Np increases behind the shock front. This can be explained both by the difficulty of ion diffusion across the magnetic field and by the changes in the mechanisms of nonlinear twisting of the shock ramp. Studies in the literature usually describe the results of simulations performed for ions reflected from the front, but not for those that have passed beyond the ramp. However, a recent paper by Ofman et al. [32] demonstrated the possibility of a strong increase in the helium abundance Nα/Np behind the shock front. The data obtained in the present study are consistent with the results of modeling reported by these authors. Ofman et al. [32] also showed that the structure of the density of He++ ions is consistent with the surfing acceleration mechanism and can produce oscillations with significant amplitude in the downstream flow.

The accumulation of helium is found to occur in both IP shock crossings and Earth's bow shock crossings, and in both cases the process similarly depends on the theta angle. Thus, the observations may suggest that an identical physical mechanism is occurring in both types of phenomenon crossings. The difference in absolute values is most likely due to the difference in the ion composition.

The problem of changes in the value of Nα/Np in the case of shocks with a large magnetosonic Mach number remains open, as the number of events suitable for consideration is insufficient. This will be also considered in future work with the help of advanced statistics, and will include the Earth's bow shock crossings.

## 5. Conclusions

Based on SPEKTR-R/BMSW plasma measurements with a high time resolution of ~ 1.5 s, the dynamics of He++ ions during satellite crossings of collisionless shocks between 2011 and 2017 was studied in detail. All of the main parameters of the He++ ions (velocity Vα, temperature Tα, absolute Nα and helium abundance Nα/Np) for 20 IP shocks and 25 Earth's bow shock crossings were analyzed using data from the BMSW instrument and measurements by the WIND, Cluster and THEMIS satellites. It was shown that the average value of the helium abundance Nα/Np behind the IP shock front is slightly less (~9%) than in the undisturbed region, and that the maximum value of this parameter is even lower behind the IP shock front. It should be noted that this does not apply to the absolute value of the He++ ion density.

For a more detailed consideration of the time profiles of Nα/Np, the dependence of this value on the parameters $\beta_i$, $M_{MS}$, and angle $\theta_{Bn}$ was analyzed. The preliminary results showed no obvious dependence of the change in Nα/Np on the parameters $\beta_i$ and $M_{MS}$. A correlation was found to exist between Nα/Np and the angle $\theta_{Bn}$: the lower the value of $\theta_{Bn}$, the greater the reduction in the abundance of helium Nα/Np behind the IP shock front. For the Earth's bow shock crossings, a significant increase in Nα/Np was found in quasi-perpendicular events. Our results correspond with those obtained from the model developed by Ofman et al. [32]. To further validate the reliability of these results, it will be necessary to continue with experimental studies of the dynamics of the small ionic components of the solar wind at shocks, and to model these using MHD and kinetic approximations. This work will form the next step in our research.

**Funding:** This research was funded by Russian Science Foundation, grant number 22-12-00227

**Data Availability Statement:** The authors are grateful to the developers of the CDAWEB database (https://cdaweb.gsfc.nasa.gov/istp_public/ (accessed on 25 August 2022)); BMSW data available by the request (http://aurora.troja.mff.cuni.cz/spektr-r/project/index.php (accessed on 25 August 2022)).

**Acknowledgments:** The authors would like to express their gratitude to NASA CDAWEB for the use of data on the plasma and magnetic field parameters measured by the WIND, THEMIS and Cluster satellites.

**Conflicts of Interest:** The authors declare no conflict of interest.